\documentclass[nofootinbib]{revtex4}
\usepackage{graphicx}
\usepackage{fancyhdr}
\usepackage{amsmath}
\usepackage{color}

\begin{document}

\title{Higgs inflation scenario in a radiative seesaw model and its testability at the ILC
\footnote{This proceeding paper is based on Ref.~\cite{Kanemura:2012ha}, with including some of the recent developments. }
}

%

\author{Toshinori Matsui}
\affiliation{Department of Physics, University of Toyama, Toyama 930-8555, Japan}

\begin{abstract}
The Higgs inflation scenario is an approach to realize the cosmic inflation, 
where the Higgs boson plays a role of the inflaton. In the minimal model, 
it would be difficult to satisfy theoretical constraints from vacuum stability and perturbative unitarity. 
These problems can be solved by considering multi-Higgs models. 
In this talk, 
we discuss a Higgs inflation scenario in a radiative seesaw model with an inert doublet, 
which originally has been proposed to explain dark matter and neutrino masses. 
We study this model under the constraints from the current data, 
and find parameter regions where additional scalar bosons can play a role of inflatons. 
They satisfy the current data from neutrino experiments, 
the dark matter searches and also from LEP and LHC. 
A unique phenomenological prediction appears in the mass spectrum of inert scalar bosons.  
We show that this scenario is challenging to be tested at the LHC, 
but would be well testable at the International Liner Collider by measuring endpoints of energy distribution of a two jet system from decay processes of the inert scalar fields produced via pair production. 
\end{abstract}

\maketitle

\thispagestyle{fancy}


\section{Introduction}

The standard cosmology is a very successful model to explain the expansion of the Universe, the abundances of the light elements and the cosmic microwave background. 
However, we need inflation to solve horizon problem and flatness problem. 
In general, the inflation is explained by the exponential expansion~\cite{inf}.
But, we do not know the detail of the inflation. 
The scenario of slow-roll inflation~\cite{slow-roll} can be realized by a scalar particle, so-called the inflaton. 
If the inflation potential is given, parameters for the slow-roll inflation can be calculated. 

We consider one possibility of inflation scenarios, the Higgs inflation scenario~\cite{Hinf}, where Higgs boson plays a role of inflaton. 
In this model, we introduce the coupling term of the Higgs field $\Phi$ with gravity as $\xi \Phi^\dagger \Phi {\mathcal R}$ (${\mathcal R}$ is the Ricci scalar). 
Then, its coupling is too large $\xi\simeq 10^{5}$ from the primordial power spectrum of the curvature perturbation . 
Slow-roll parameters which are calculated by the inflation potential must satisfy the data from the Planck experiment~\cite{Ade:2013uln}. 
The inflation scale ($\Lambda_{\rm I} = \frac{M_P}{\sqrt{\xi}}$ for the Higgs inflation scenario) is also calculated from the inflation potential.
Constraints of the slow-roll inflation scenario can be satisfied with experiments. 
Especially, the data from the Planck experiment~\cite{Ade:2013uln} support the Higgs inflation scenario\footnote{
New result of the B-mode polarization~\cite{Ade:2014xna} shows the high inflation scale. 
To reconcile the Higgs inflation models with the BICEP2 data, we must improve models of the Higgs inflation. 
The paper~\cite{Nakayama:2014koa} shows that the scenario of the standard model Higgs field as a inflaton is possible if we modify the kinetic term at large field value. 
The other Higgs inflation models~\cite{Hinf_flat} explain this experimental result by the flat Higgs potential around the Planck scale. 
In this talk, we do not consider this experimental result.
}. 

However, there are some theoretical problems in the simplest model. 
When we calculate the running coupling constant of the Higgs self-coupling, 
the critical energy scale is around $10^{10}$~GeV due to the contribution of the top quark~\cite{lambda_run}. 
The vacuum is difficult to be stable up to the inflation scale $\Lambda_{\rm I}$. 
This problem can be solved in two Higgs doublet models~\cite{Deshpande:1977rw}. 
Because the loop effect of additional scalar bosons weakens the top-loop contribution in the running coupling constants~\cite{extended_run}.
Perturbative unitarity is also violated at the energy scale $\Lambda_{\rm U} = \frac{M_P}{\xi}$ by the Higgs-gauge scattering processes~\cite{uni_break}.
This problem is solved by a heavy additional real singlet scalar boson which does not interact with gauge fields as shown by~\cite{uni_care}. 

In this talk, we explain not only dark matter, neutrino masses but also inflation. 
We show a radiative seesaw scenario with the multi-Higgs structure, which was proposed by E. Ma~\cite{Ma}, is constrained by the inflation condition. 
We discuss the testability of the characteristic mass spectrum at the collider experiments.

\section{The radiative seesaw model}
\begin{figure}[t]
  \begin{center}
    \begin{tabular}{c}
      \begin{minipage}[b]{0.5\hsize}
        \begin{center}
        \makeatletter
\def\@captype{table}
\makeatother
  \begin{tabular}
{|p{12mm}|@{\vrule width 1.8pt\ }p{6mm}|p{6mm}|p{6mm}|p{6mm}|p{6mm}|p{6mm}|p{6mm}|p{6mm}|p{6mm}|}
   \hline
     &$Q_L$&$u_R$&$d_R$&$L_L$&$\ell_R$&$\Phi_1$&$\Phi_2$&$\nu_R^{}$&$\sigma$
    \\ \noalign{\hrule height 1.8pt}
    SU(3)$_{\rm C}$&{\bf 3}&{\bf 3}&{\bf 3}&{\bf 1}&{\bf 1}&{\bf 1}&{\bf 1}&{\bf 1}&{\bf 1}
    \\ \hline
    SU(2)$_{\rm I}$&{\bf 2}&{\bf 1}&{\bf 1}&{\bf 2}&{\bf 1}&{\bf 2}&{\bf 2}&{\bf 1}&{\bf 1}
    \\ \hline
    U(1)$_{\rm Y}$&$\frac{1}{6}$&$\frac{2}{3}$&$-\frac{1}{3}$&$-\frac{1}{2}$&$-$1&$\frac{1}{2}$&$\frac{1}{2}$&0&0
    \\ \hline
    $Z_2$&1&1&1&1&1&1&$-$1&$-$1&$-$1
    \\ \hline
\end{tabular}
\caption{Particle contents and their quantum charges.}
\label{table:particle}
        \end{center}
      \end{minipage}
\quad
      \begin{minipage}[b]{0.5\hsize}
        \begin{center}
    \includegraphics[width=5cm, clip]{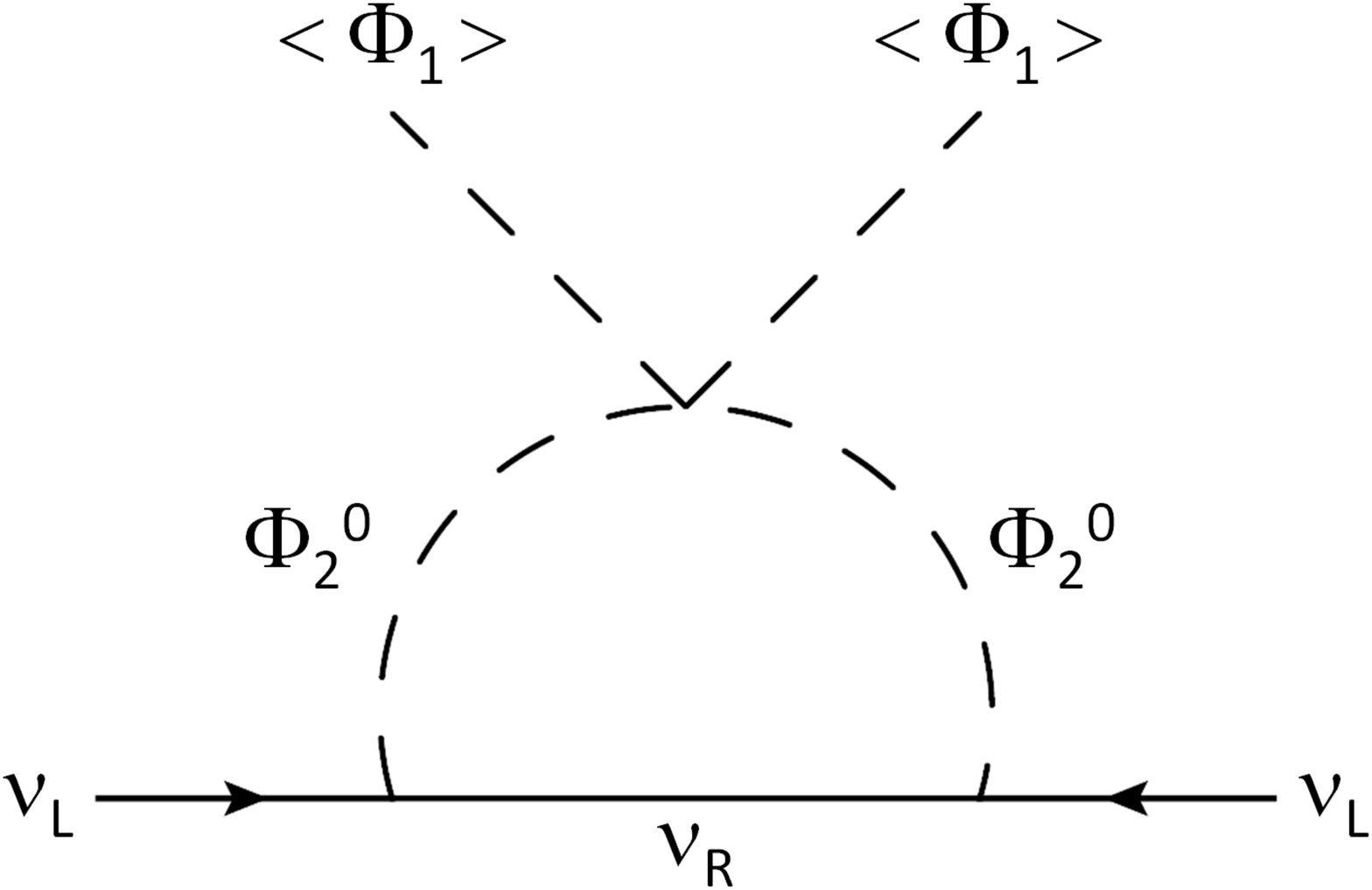}
    \caption{The Feynman diagram for tiny neutrino masses.}
    \label{fig:nmass}
          \end{center}
      \end{minipage}
    \end{tabular}
  \end{center}
\end{figure}

In our model, we introduce the second scalar doublet~$\Phi_2$, right handed neutrinos~$\nu_R^{\,i}$ ($i=1-3$) and real singlet scalar~$\sigma$ and impose quantum numbers under the an unbroken discrete $Z_2$ symmetry shown in Table~\ref{table:particle}. 
%

The Yukawa interaction for leptons and the Majorana mass term are given by
\begin{eqnarray}
{\cal L}_{\rm Yukawa} = Y_\ell \overline{L_L}\Phi_1\ell_R+Y_\nu\overline{L_L}\Phi_2^c\nu_R+h.c. \, , \qquad \qquad
{\cal L}_{\rm Majorana} = \frac{1}{2}M_R^k\overline{(\nu_R^k)^c}\nu_R^k, 
\end{eqnarray}
where the superscript $c$ denotes the charge conjugation. 
In the Feynman diagram in Fig.~\ref{fig:nmass}, which is explained by Ref.~\cite{Ma}, 
the extra lightest neutral particle can be a dark matter candidate by $Z_2$ symmetry. 
We can explain neutrino masses at the loop level by
\begin{eqnarray}
(m_{\nu})_{ij}
=
\sum_{k}
\frac{(Y_\nu)_i^k(Y_\nu)_j^kM_R^k}{16\pi^2}
\left[\frac{m_H^2}{m_H^2-\left(M_R^k\right)^2}\ln\frac{m_H^2}{\left(M_R^k\right)^2}
-\frac{m_A^2}{m_A^2-\left(M_R^k\right)^2}\ln\frac{m_A^2}{\left(M_R^k\right)^2}\right].  
\end{eqnarray}
%
The neutrino oscillation data is explained by neutrino Yukawa coupling constants $(Y_\nu)_i^k$, which satisfy $(Y_\nu)_i^k(Y_\nu)_j^k/M_R^k\simeq {\cal O}(10^{-7})$~GeV$^{-1}$. 

The Higgs potential is given by
\begin{eqnarray}
V_{\rm J}
&=&
\frac{1}{2}\left( 1+ \frac{2\xi_1|\Phi_1|^2+2\xi_2|\Phi_2|^2+\zeta \sigma^2}{M_P^2} \right) M_P^2{\mathcal R} \notag \\
&+&
\mu_1^2|\Phi_1|^2+\mu_2^2|\Phi_2|^2+\mu_\sigma^2\sigma^2+\mu_{\sigma \phi}[\sigma(\Phi_1^\dagger \Phi_2)^2+h.c.] 
+ \frac{1}{2}\lambda_1|\Phi_1|^4+\frac{1}{2}\lambda_2|\Phi_2|^4+\lambda_\sigma\sigma^4 \notag \\ 
&+&
\lambda_3|\Phi_1|^2|\Phi_2|^2+\lambda_4(\Phi_1^\dagger\Phi_2)(\Phi_2^\dagger\Phi_1)+\frac{1}{2}\lambda_5[(\Phi_1^\dagger \Phi_2)^2+h.c.] 
+ \lambda_{\sigma 1}|\Phi_1|^2 \sigma^2 +  \lambda_{\sigma 2}|\Phi_2|^2 \sigma^2.
\end{eqnarray}
%
When we assume $\mu_1^2 < $0 and $\mu_2^2 >$ 0, $\Phi_1$ obtains the vacuum expectation value (VEV) $v$ ($=\sqrt{-2\mu_1^2/\lambda_1}$), while $\Phi_2$, which has the odd-quantum number of the $Z_2$ symmetry, cannot get the VEV. 
Mass eigenstates of the scalar bosons are the SM-like $Z_2$-even Higgs scalar boson~$h$, the $Z_2$-odd CP-even scalar boson~$H$, the $Z_2$-odd CP-odd scalar boson~$A$ and $Z_2$-odd charged scalar bosons~$H^\pm$. 
Masses of these scalar bosons are given in Ref.~\cite{Ma}; $m_h^2=\lambda_1 v^2, \ m_H^2=\mu_2^2 +\frac{1}{2}(\lambda_3+\lambda_4+\lambda_5) v^2, \ m_A^2=\mu_2^2 +\frac{1}{2}(\lambda_3+\lambda_4-\lambda_5) v^2, \ m_{H^{\pm}}^2=\mu_2^2 +\frac{1}{2}\lambda_3 v^2$. 
As the $Z_2$-odd neutral singlet scalar~$\sigma$ is constrained by perturbative unitarity~\cite{Kang:2013zba}: $m_\sigma \leq \Lambda_U$, we assume that $m_\sigma$ is heavy enough, so that it gives an insignificant effect on phenomenology. 
For simplicity, we take $\mu_{\sigma \phi}=\lambda_{\sigma 1}=\lambda_{\sigma 2}=0$ and $\xi_1\simeq \xi_2 \ll \zeta$. 
We study parameter regions which satisfy the conditions of vacuum stability and perturbative unitarity. 

\section{Constraints on the parameters}

The Higgs potential in the Einstein frame is given by 
\begin{eqnarray}
V_{\rm E}=\frac{V_{\rm J}}{\Omega^4}
= \frac{M_P^4}{8} \frac{\lambda_1h_1^4+\lambda_2h_2^4+\lambda_\sigma\sigma^4+2\{\lambda_3+\lambda_4+\lambda_5\cos(2\theta)\} h_1^2h_2^2 + \lambda_{\sigma 1} h_1^2 \sigma^2 + \lambda_{\sigma 2} h_2^2 \sigma^2}
{(M_P^2 + \xi_1h_1^2+\xi_2h_2^2+\zeta \sigma^2)^2}, 
\end{eqnarray}
where
\begin{eqnarray}
\Omega^2=1+ \frac{2\xi_1|\Phi_1|^2+2\xi_2|\Phi_2|^2+\zeta \sigma^2}{M_P^2},\,\Phi_1=\begin{pmatrix} 0 \\ h_1 \end{pmatrix},\,\Phi_2=\begin{pmatrix} 0 \\ h_2 e^{i\theta} \end{pmatrix}. 
\end{eqnarray}
%
For small field values~$\Omega^2\simeq 1$, the potential is the same as Jordan frame for the initial Higgs field~($V_{\rm E}\simeq V_{\rm J}$). 
On the other hand, for large fields values~$\Omega^2 \gg 1$, we define 
$\varphi \equiv \sqrt{\frac{3}{2}} M_P \ln \Omega^2, \,\, 
r_2 \equiv \frac{h_2}{h_1}, \,\, r_{\sigma} \equiv \frac{\sigma}{h_1}$. 
For stabilizing $r_2$, $r_\sigma$ as a finite value, we need to impose following condition: 
\begin{eqnarray}
\lambda_1\lambda_2-(\lambda_3+\lambda_4)^2&>&0.
 \label{eq:const_inf}
\end{eqnarray}
%
This is the constraint from the inflation on our model because the heavy particle $\sigma$ dominantly plays a role of inflaton.

The CP-odd boson $A$ is assumed to be the lightest $Z_2$-odd particle; i.e., the dark matter candidate. 
When we change the sign of the coupling constant $\lambda_5$, 
the similar discussion can be applied for the case of the CP-even boson $H$ to be the lightest. 
As $\lambda_5$ can be sizable which is not constrained from the inflation, 
the dominant scattering process is $AN\to AN$ ($N$ is a nucleon) where the standard model-like Higgs boson is propagating. 
We can avoid the process $AN\to HN$ kinematically, and the cross section is consistent with the current direct search results for dark matter. 
As shown in~\cite{LopezHonorez:2006gr,dark matter_direct}, 
the cross section of $AN\to AN$ process is
\begin{eqnarray}
\sigma(AN\to AN)\simeq\frac{\lambda_{hAA}^2}{4m_h^4}\frac{m_N^2}{\pi(m_A+m_N)^2}f_N^2,
\end{eqnarray}
where
$\lambda_{hAA} \equiv \lambda_3+\lambda_4-\lambda_5$, 
$f_N \equiv \sum_{q}m_Nf_{Tq}+\frac{2}{9}m_Nf_{TG}$ and 
$m_N$ is the mass of nucleon, where 
$f_{Tu}+f_{Td}=0.056$, $f_{Ts}=0$~\cite{lattice} and $f_{TG}=0.944$~\cite{trace-anomaly}.
To satisfy the data of the dark matter relic abundance from the Planck experiment~\cite{Ade:2013uln} and the data of the upper bound on the scattering cross section for $AN\to AN$ from the  experiments $\sigma \simeq 2\times 10^{-45} {\rm cm}^{2}$~\cite{XENON100, Akerib:2013tjd},  
the coupling constant $\lambda_{hAA}$ is required to satisfy 
\begin{eqnarray}
\lambda_{hAA}\lesssim 0.036,
\label{eq:hAA}
\end{eqnarray}
at the electroweak scale. 
When $\lambda_5$ is not small, the co-annihilation process $AH\to XX$ via the $Z$ boson does not contribute to the dark matter relic abundance. 
This case is the same as the singlet scalar dark matter model~\cite{Djouadi:2011aa, Cline:2013gha}. 
On the other hand, to avoid the current invisible decay $h \to AA$ kinematically~\cite{inertdark matter, Espinosa:2012im}, $m_A$ must be bigger than $m_h/2$. 
To satisfy these dark matter conditions, we require 
\begin{eqnarray}
63~{\rm GeV} \lesssim m_A \lesssim 66~{\rm GeV}.
\label{eq:dark matter_mass}
\end{eqnarray}

%
\begin{figure}[t]
  \begin{center}
    \begin{tabular}{c}
\begin{minipage}[b]{0.497\textwidth}
 \begin{center}
    \includegraphics[width=9cm, clip]{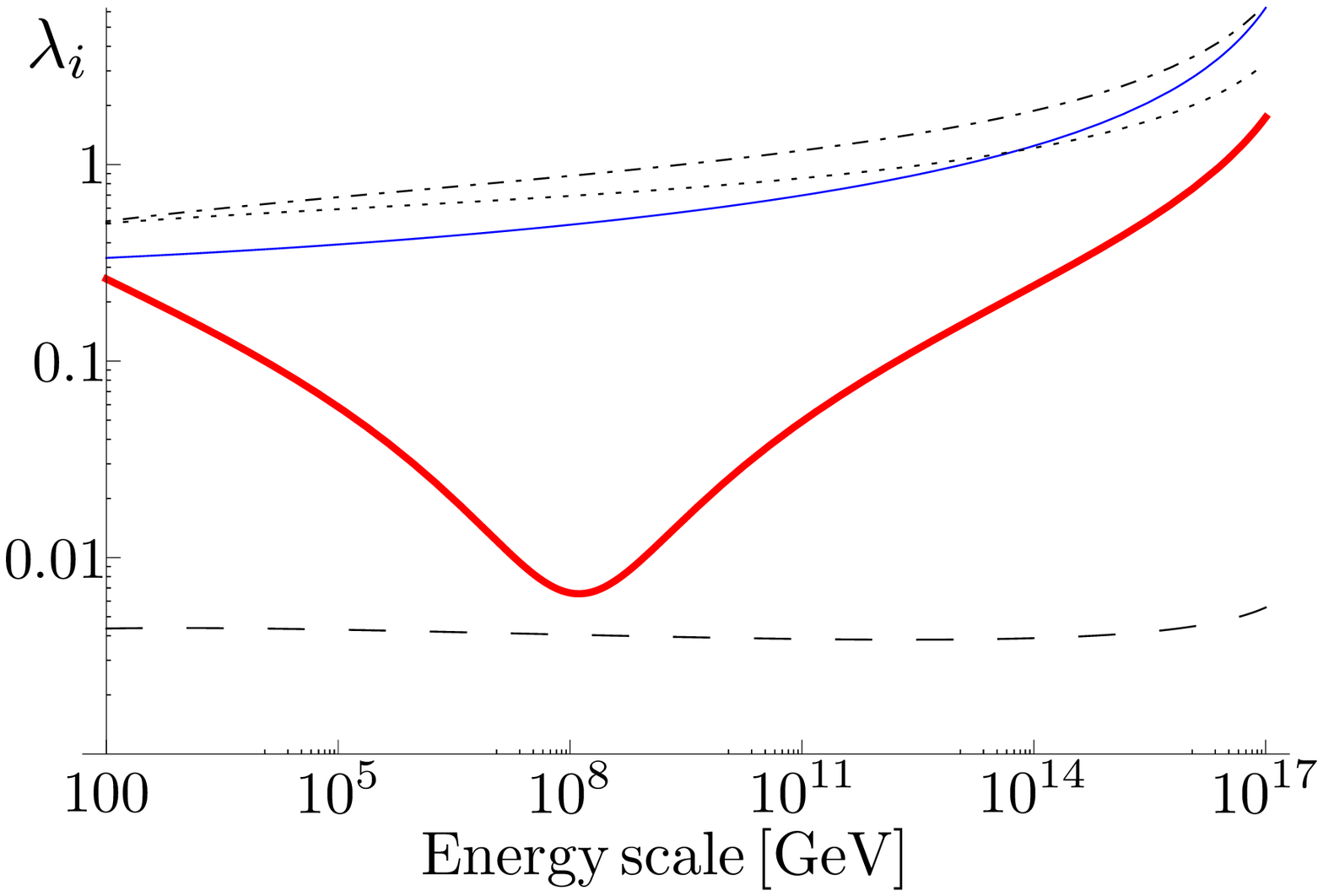}
      \caption{Running of the scalar coupling constants. 
  Red (solid), blue (dashed), brown (dot-dashed), green (dotted) and black (long-dashed) curves show $\lambda_1$, $\lambda_2$, $\lambda_3$, $|\lambda_4|$ and $\lambda_5$, respectively. }
    \label{fig:running}
     \end{center}
  \end{minipage}
\quad
 \begin{minipage}[b]{0.497\textwidth}
\begin{center}
\makeatletter
\def\@captype{table}
\makeatother
\begin{tabular}{|p{15mm}|@{\vrule width 1.8pt\ }p{10mm}|p{10mm}|p{10mm}|p{10mm}|p{16mm}|}
\hline
&$\lambda_{1}$&$\lambda_{2}$&$\lambda_{3}$&$\lambda_{4}$&$\lambda_{5}$ \\ \noalign{\hrule height 1.8pt}
 $10^{2}$~GeV
  &0.262 &0.335&0.514&$-0.503$&4.35$\times10^{-3}$\\
\hline
 $10^{17}$~GeV
  &1.74 &6.28&6.60&$-3.30$&5.57$\times10^{-3}$\\
\hline
\end{tabular}
\caption{The possible parameter set which satisfies constraints from the inflation condition and the dark matter data at the scales of $10^{2}$~GeV and $10^{17}$~GeV. \\}
\label{table:lambda}
\end{center} 
  \end{minipage}
      \end{tabular}
  \end{center}
\end{figure}

%
Take into account the above conditions, 
the vacuum stability condition
\begin{eqnarray}
\lambda_1>0,\ \lambda_2>0,\ \sqrt{\lambda_1 \lambda_2}+\lambda_3+{\rm min}[0, \lambda_4+\lambda_5, \lambda_4-\lambda_5]>0, 
\end{eqnarray}
and the conditions of triviality $\lambda_i \lesssim 2\pi$, 
we analyze the renormalization group equations~\cite{beta}. 
In Fig.~\ref{fig:running}, 
running of the scalar coupling constants is shown 
between the electroweak scale and the inflation scale. 
In Table~\ref{table:lambda}, 
we show the values of the scalar coupling constants 
at the scales of ${\cal O}(10^{2})$~GeV and ${\cal O}(10^{17})$~GeV, 
which satisfy the conditions of the dark matter and the inflation. 
From this parameter set, mass spectrum of the scalar bosons is constrained by 
\begin{eqnarray}
m_H\lesssim100~{\rm GeV}, 142~{\rm GeV}\lesssim m_{H^\pm}\lesssim146~{\rm GeV}. 
\label{eq:scalar_mass}
\end{eqnarray}

\section{Collider Phenomenology}
%
In this scenario, $m_{H^\pm}$ is about 140~GeV. 
This value satisfies the lower bound from the LEP experiment~\cite{LEP_direct,LEP_pm}. 
From the measurement of the $Z$ boson decay width, $m_H+m_A$ is greater than $m_Z$~\cite{LEP_direct,LEP_HApair}. 
Moreover, the direct detection of dark matter at LEP give a constraint on $HA$ pair production~\cite{LEP_HApair}. 
Because of the constraint from the inflation $m_H\lesssim100~{\rm GeV}$, the mass difference between the two inert scalar bosons is allowed only in a narrow region~\cite{LEP_direct,LEP_HApair}: 
\begin{eqnarray}
m_H-m_A<8~{\rm GeV}. 
\label{dark matter_mass}
\end{eqnarray}

\begin{figure}[b]
  \begin{center}
    \begin{tabular}{c}
      \begin{minipage}[b]{0.45\hsize}
        \begin{center}
          \includegraphics[clip, width=5.9cm]{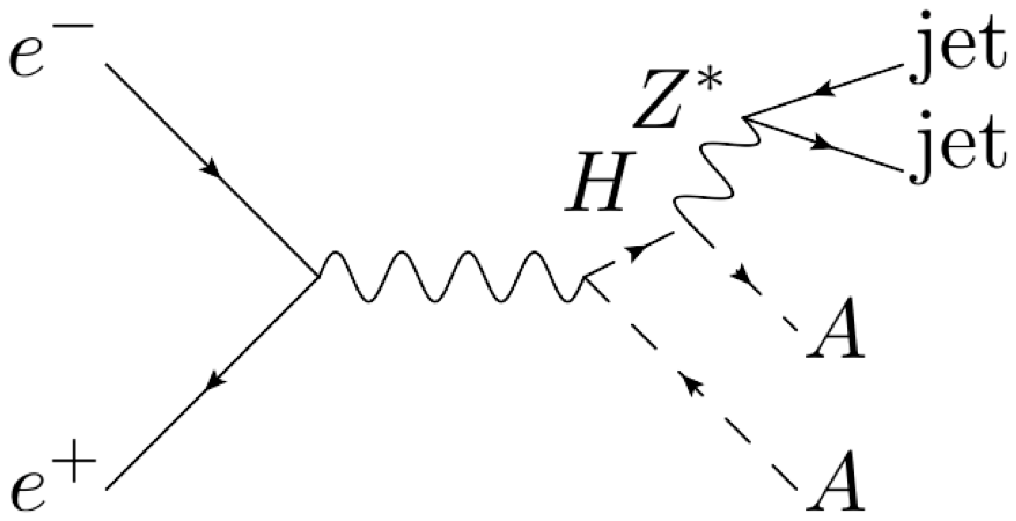}
              \caption{The signal of $HA$ production at the ILC. \\}
              \label{fig:diagram_even}
        \end{center}
      \end{minipage}
 \hspace{1.5pc}
      \begin{minipage}[b]{0.45\hsize}
        \begin{center}
          \includegraphics[clip, width=5.9cm]{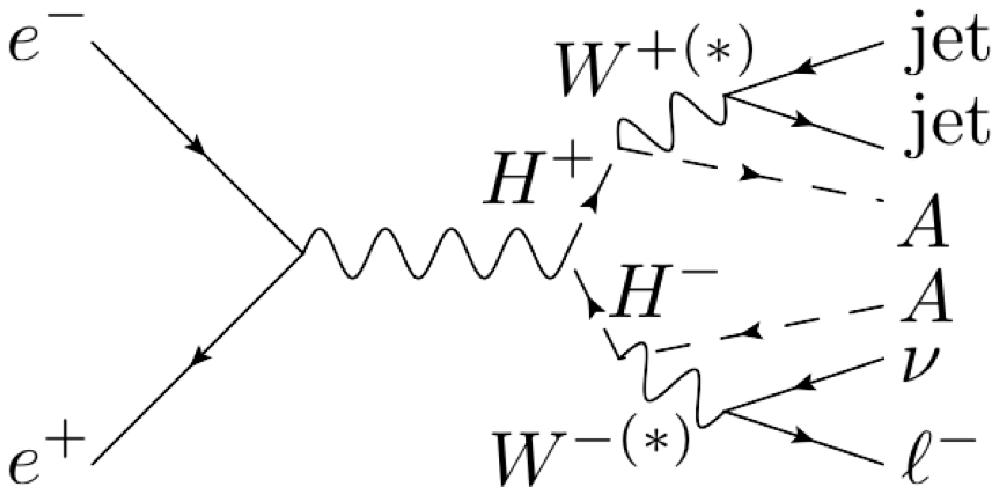}
              \caption{The signal of $H^+H^-$ production at the ILC. }
              \label{fig:diagram_charged}
        \end{center}
      \end{minipage}
    \end{tabular}
  \end{center}
\end{figure}
%
\begin{figure}[t]
  \begin{center}
    \begin{tabular}{c}
      \begin{minipage}[b]{0.45\hsize}
        \begin{center}
          \includegraphics[clip, width=8cm]{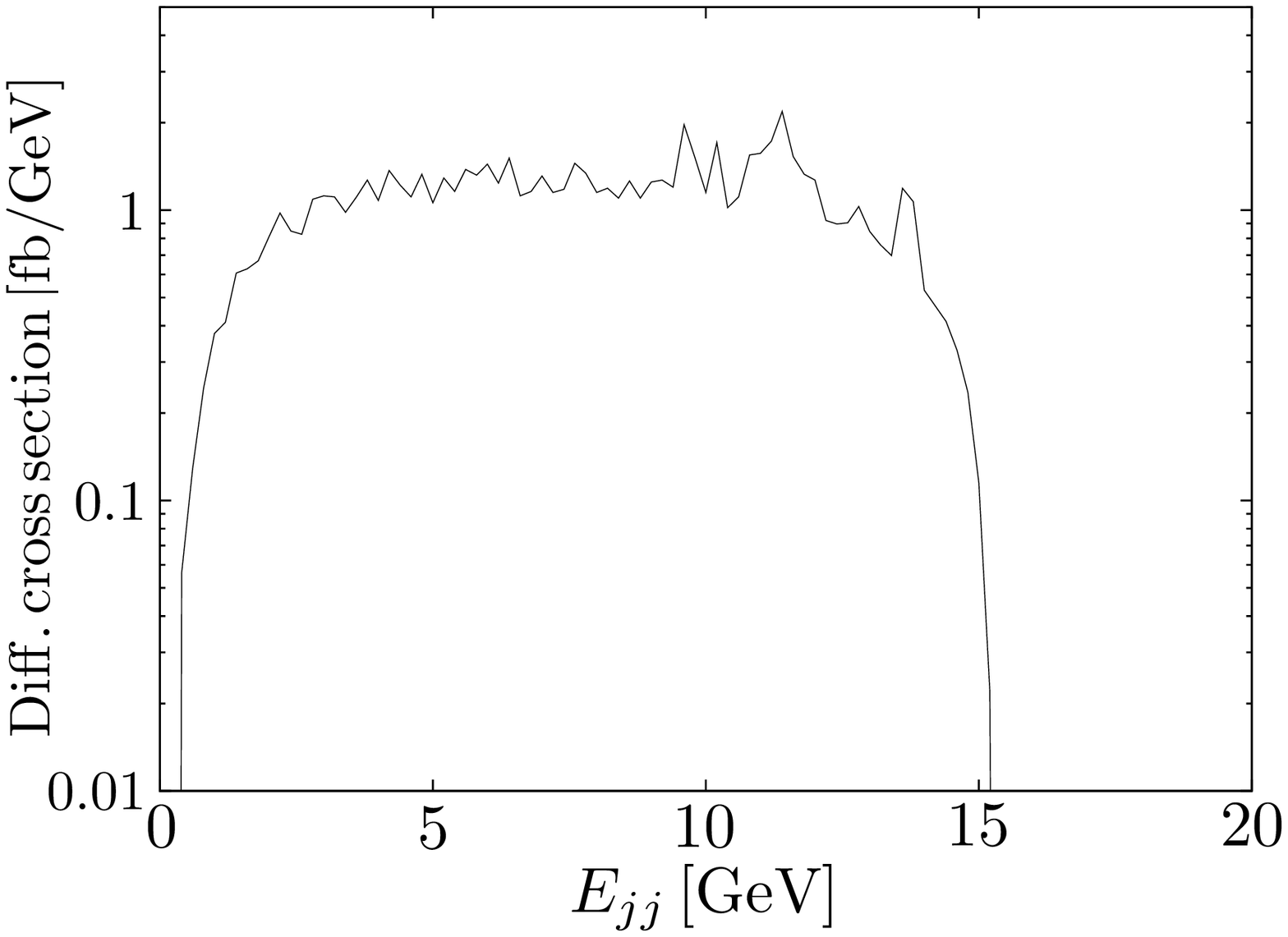}
              \caption{The distribution of $E_{jj}$ for the cross section for
$e^+e^-\to HA\to AAZ^*\to AAjj$. 
In our parameter set, the endpoint in the $E_{jj}$ distribution is estimated at $0.28~{\rm GeV} < E_{jj} < 15~{\rm GeV}$. 
This value corresponds to $m_H=67$~GeV, $m_A=65$~GeV. }
              \label{fig:distribution_even}
        \end{center}
      \end{minipage}
        \hspace{1.5pc}
      \begin{minipage}[b]{0.47\hsize}
        \begin{center}
          \includegraphics[clip, width=8cm]{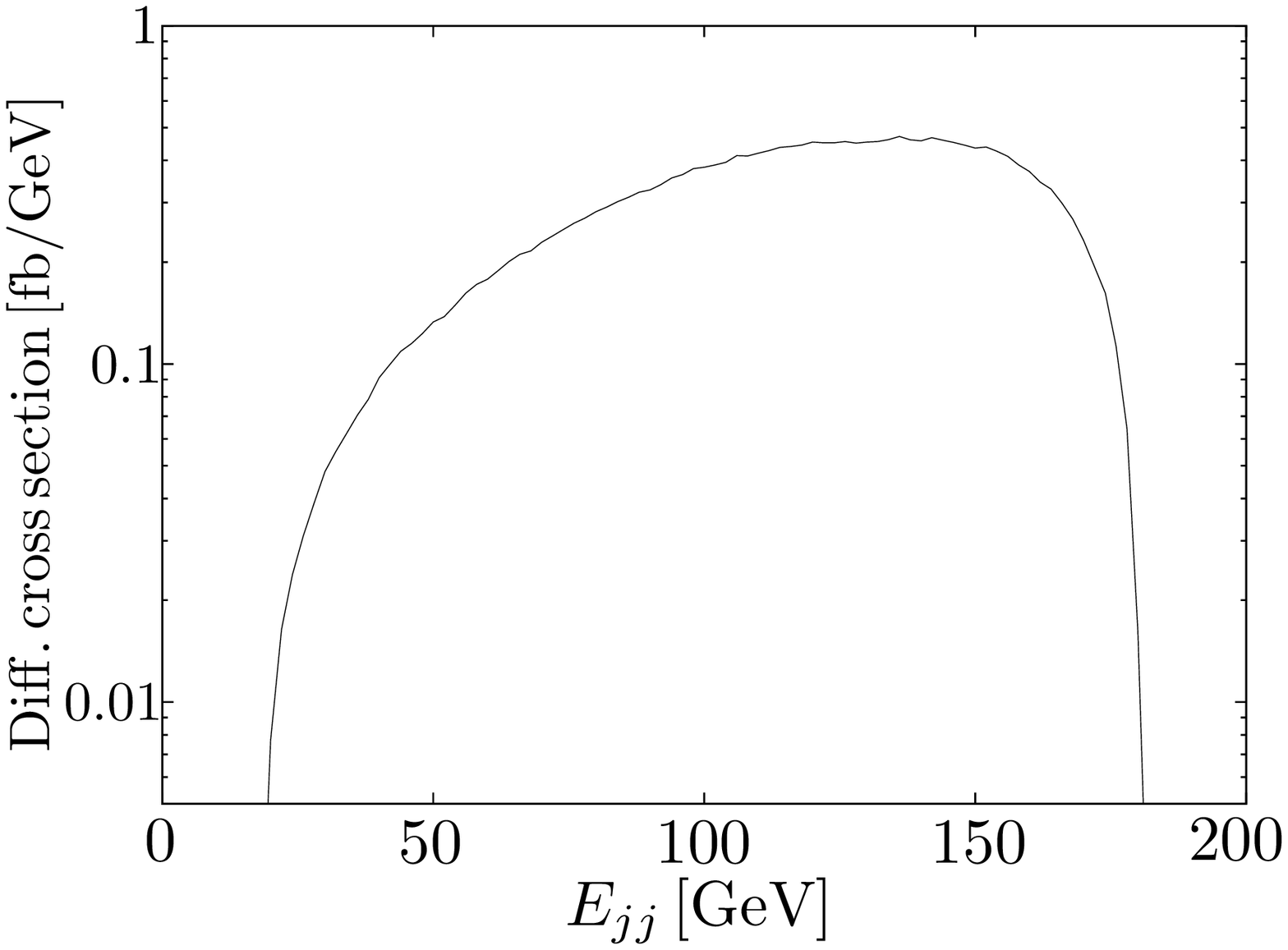}
              \caption{The distribution of $E_{jj}$ for the cross section for 
  $e^+e^-\to H^+H^-\to W^{+(*)}W^{-(*)}AA\to jj\ell\nu AA$. 
  In our parameter set, the endpoint in the $E_{jj}$ distribution is estimated at $17~{\rm GeV} < E_{jj} < 180~{\rm GeV}$.
  This value corresponds to $m_{H^\pm}=140$~GeV, $m_A=65$~GeV. }
              \label{fig:distribution_charged}
        \end{center}
      \end{minipage}
    \end{tabular}
  \end{center}
\end{figure}

%
In Ref.~\cite{Dolle:2009ft}, the collider phenomenology in the inert doublet model is discussed at the LHC with $\sqrt{s}=14$~TeV. 
According to their work, the process of $q\overline{q}\to Z \to HA \to Z^{(*)}AA \to \ell^+\ell^- AA$ is dominant.
They chose the mass difference of inert neutral scalar bosons to be 10, 50 and 70~GeV. 
As $m_A$ is 65~GeV in our model, 
if the mass difference becomes large, 
inflation condition Eq.~(\ref{eq:const_inf}) cannot be satisfied. 
On the contrary, 
if the mass difference become small, the signal is also small ($S/\sqrt{B}=0.02$). 
Therefore, the model is difficult to be tested at the LHC.

Let us discuss the signals of $H, A$ and $H^\pm$ at the ILC with $\sqrt{s}=500$~GeV. 
In this analysis, we use Calchep~2.5.6 for numerical evaluation~\cite{calc}.
The detail which contains background analysis of inert doublet model is disused in the paper~\cite{inILC} which is applicable to our model. 
First, the dominant signal of the $HA$ production process is  $e^+e^- \rightarrow Z^* \rightarrow HA \rightarrow AAZ^*\rightarrow AAjj$ ($j$: jet, $Z^*$: off-shell $Z$ boson) shown in Fig.~\ref{fig:diagram_even}. 
The final state is two jets with a missing momentum.
The energy of the two-jet system $E_{jj}$ satisfies the following equation because of the kinematical constraint given as 
\begin{eqnarray}
\frac{m_H^2-m_A^2}{4m_H^2} \left(\sqrt{s}-\sqrt{s-4m_H^2} \right)<E_{jj}<\frac{m_H^2-m_A^2}{4m_H^2} \left(\sqrt{s}+\sqrt{s-4m_H^2} \right).
\end{eqnarray}
%
When the center of mass energy is $\sqrt{s}=500$~GeV, $E_{jj}$ is evaluated by using our parameter set as $0.28~{\rm GeV}<E_{jj}<15~{\rm GeV}$. 
The distribution of $E_{jj}$ of the cross section for this prosecc is shown in Fig.~\ref{fig:distribution_even}.
We expect that $m_H$ and $m_A$ can be measured by using the endpoints in the $E_{jj}$ distribution at the ILC after the background reduction. 

Next, the dominant signal of the $H^+ H^-$ production process is $e^+e^- \rightarrow Z^* (\gamma^*) \rightarrow H^+H^- \rightarrow W^{+(*)}W^{-(*)}AA \rightarrow jjl\nu AA$ ($W^{\pm(*)}$ is off-shell $W$ boson) as shown in Fig.~\ref{fig:diagram_charged}. 
The final state of this process is a charged lepton and two jets with the missing momentum. 
From the same discussion, the energy of the two-jet system, $E_{jj}$ is constrained as 
\begin{eqnarray}
\frac{m_{H^\pm}^2-m_A^2}{4m_{H^\pm}^2} \left(\sqrt{s}-\sqrt{s-4m_{H^\pm}} \right)<E_{jj}<\frac{m_{H^\pm}^2-m_A^2}{4m_{H^\pm}^2} \left(\sqrt{s}+\sqrt{s-4m_{H^\pm}} \right).
\end{eqnarray}
%
When the center of mass energy is $\sqrt{s}=500$~GeV, $E_{jj}$ is evaluated by using our parameter set as $17~{\rm GeV}<E_{jj}<180~{\rm GeV}$. 
The distribution of $E_{jj}$ of the cross section for this process is shown in Fig.~\ref{fig:distribution_charged}.
We expect that $m_{H^\pm}$ and $m_A$ can be measured by using the endpoints in the $E_{jj}$ distribution at the ILC after the background reduction. 
Backgrounds could also be reduced by imposing kinematic cuts. 
We can measure $m_{H^\pm}$ and $m_A$ by observing the endpoints in the $E_{jj}$ distribution at the ILC.

\section{Conclusion}
In the original Higgs inflation scenario, 
it would be difficult to satisfy perturbative unitarity and vacuum stability.
These problems can be solved by considering multi-Higgs models. 
In the framework of the radiative seesaw scenario with the multi-Higgs structure, 
we can explain not only dark matter, 
neutrino masses but also inflation. 
This scenario would be testable at the ILC by measuring the energy distribution of the inert scalar pair production.

\begin{acknowledgments}
This work is based on the collaboration with Shinya Kanemura and Takehiro Nabeshima. 
I would like to thank them for their support.
\end{acknowledgments}

\bigskip 

\end{document}